\PassOptionsToPackage{pdfpagelabels=false}{hyperref}

\documentclass[a4paper,fleqn,usenatbib]{mnras}

\usepackage{txfonts}
\usepackage[T1]{fontenc}
\usepackage{ae,aecompl}
\usepackage{graphicx}
\usepackage[english]{babel}

\title[Transient ULX in M86]{Discovery of a transient ultraluminous X-ray source in the elliptical galaxy M86}

\author[van~Haaften et al.]{Lennart~M.~van~Haaften,$^{1}$\thanks{E-mail: L.vanHaaften@ttu.edu}
Thomas~J.~Maccarone,$^{1}$
Katherine~L.~Rhode,$^{2}$
\newauthor Arunav~Kundu,$^{1,3,4}$
and Stephen~E.~Zepf\,$^{5}$
\\
$^{1}$Department of Physics and Astronomy, Texas Tech University, Box 41051, Lubbock, TX 79409-1051, USA \\
$^{2}$Department of Astronomy, Indiana University, 727 East Third Street, Bloomington, IN 47405, USA \\
$^{3}$Eureka Scientific, Inc., 2452 Delmer Street, Suite 100, Oakland, CA 94602, USA \\
$^{4}$Computational Physics, Inc., 8001 Braddock Road, Suite 210, Springfield, VA 22151, USA \\
$^{5}$Department of Physics and Astronomy, Michigan State University, East Lansing, MI 48824, USA
}

\date{Accepted 2018 November 23. Received 2018 November 20; in original form 2018 July 20}

\pubyear{2018}

\begin{document}
\label{firstpage}
\pagerange{\pageref{firstpage}--\pageref{lastpage}}
\maketitle

\begin{abstract}
We report the discovery of the transient ultraluminous X-ray source (ULX) CXOU J122602.3+125951 (hereafter M86 tULX-1), located $3\arcmin 52\arcsec$ ($19$ kpc) northwest of the centre of the giant elliptical galaxy M86 (NGC 4406) in the Virgo Cluster.
The spectrum of M86 tULX-1 can be fit by a power law plus multicolour disc model with a $1.0^{+0.8}_{-2.6}$ index and a $0.66^{+0.17}_{-0.11}$ keV inner disc temperature, or by a power law with a $1.86 \pm 0.10$ index.
For an isotropically emitting source at the distance of M86, the luminosity based on the superposition of spectral models is ${(5 \pm 1) \times 10^{39}}$ erg s$^{-1}$.
Its relatively hard spectrum places M86 tULX-1 in a hitherto unpopulated region in the luminosity-disc temperature diagram, between other ULXs and the (sub-Eddington) black-hole X-ray binaries.
We discovered M86 tULX-1 in an archival 148-ks 2013 July \textit{Chandra} observation, and it was not detected in a 20-ks 2016 May \textit{Chandra} observation, meaning it faded by a factor of at least $30$ in three years.
Based on our analysis of deep optical imaging of M86, it is probably not located in a globular cluster. It is the brightest ULX found in an old field environment unaffected by recent galaxy interaction.
We conclude that M86 tULX-1 may be a stellar-mass black hole of ${\sim} 30-100\ \mathrm{M}_{\odot}$ with a low-mass giant companion, or a transitional object in a state between the normal stellar-mass black holes and the ultraluminous state.
\end{abstract}


\section{Introduction}
\label{sect:ulx:intro}

Ultraluminous X-ray sources (ULXs) are point-like X-ray sources brighter than the $2 \times 10^{39}$ erg s$^{-1}$ Eddington luminosity of stellar-mass \citep[$M \lesssim 20\ \mathrm{M}_{\odot}$,][]{begelman2002,zezas2002} black holes (BHs) that are not located in galactic centres \citep[for recent reviews, see][]{bachetti2016,kaaret2017}. Most ULXs are thought to be either stellar-mass BHs, or BHs with masses in the range of $30 - 90\ \mathrm{M}_{\odot}$ \citep{zampieri2009}, or neutron stars (NSs) \citep{bachetti2014,furst2016,israel2017mnras,israel2017sci} accreting above their Eddington limits, possibly in some cases with beamed emission \citep{king2001,poutanen2007}.
One of the arguments in favour of super-Eddington accretion is that many ULX spectra show a soft excess in combination with a hard component with a high-energy curvature above ${\sim} 3$ keV, which has been interpreted as an optically thick radiatively driven wind from the outer accretion disc, pointing to an ultraluminous (UL) state \citep{roberts2007,soria2007,gladstone2009,sutton2013}.
The wind scenario is also supported by the detection of blueshifted absorption lines in the high-resolution X-ray spectra of several ULXs \citep{pinto2016,kosec2018}.
Especially ULXs with X-ray luminosity $L_\mathrm{X} = (3 - 20) \times 10^{39}$ erg s$^{-1}$ have spectra consistent with stellar-mass BHs in the UL state \citep{middleton2015}.
Alternatively, intermediate-mass black holes (IMBHs) with masses of ${\sim} 10^{2-5}\ \mathrm{M}_{\odot}$ could produce the observed X-ray luminosities with sub-Eddington accretion rates \citep{colbert1999}, and are plausible explanations for at least the ULXs with $L_\mathrm{X} \gtrsim 5 \times 10^{40}$ erg s$^{-1}$ \citep{farrell2009,davis2011,webb2012}.

Clear differences exist between the properties of ULXs in different environments.
ULXs are most commonly found in spiral galaxies \citep{swartz2004}, starburst galaxies \citep[e.g.,][]{fabbiano2001,lira2002}, and in or close to star-forming regions within galaxies \citep{swartz2009}. The association with star-forming regions favours highly magnetic NS accretors \citep{shao2015,king2016,pintore2017,koliopanos2017} and/or high-mass donor stars \citep{swartz2009,liu2009,motch2011,poutanen2013}, although it is difficult to distinguish optical emission from a companion star from accretion disc emission \citep[e.g.,][]{copperwheat2005,copperwheat2007,patruno2008,madhusudhan2008}.

Smaller numbers of ULXs are located in elliptical galaxies. In particular bright ULXs, with $L_\mathrm{X} \gtrsim 2 \times 10^{39}$ erg s$^{-1}$, are extremely rare in elliptical galaxies \citep{irwin2004,peacock2016}, although some are known in globular clusters (GCs) in ellipticals \citep{maccarone2007,irwin2010}. This suggests that bright ULXs in old populations typically have different origins and properties than equally bright ULXs in young environments, and also different from less bright ULXs in old environments, perhaps by means of having a more massive BH accretor.

In this paper we report the discovery of CXOU J122602.3+125951 (hereafter M86 tULX-1), a ULX, in the giant Virgo Cluster elliptical galaxy M86.
In Section~\ref{sect:ulx:results}, we report on this source's X-ray variability, spectrum, and luminosity, and our search for counterparts in archival optical data.
In Section~\ref{sect:ulx:discussion}, we compare this source's properties and environment to those of other bright ULXs. We also discuss possible accretor type and mass, and the companion star.
Finally, in Section~\ref{sect:ulx:alt}, we argue that M86 tULX-1 is genuinely located in M86, and not a foreground or background object.

\section{Observations and data analysis}
\label{sect:ulx:obs}

\begin{table*}
\caption{\textit{Chandra} observations of M86 used in this paper.}
\label{tab:obs}
\begin{tabular}{ccccccc}
\hline
Observation ID & Instrument & Exposure time (ks) & Data mode & RA & Dec & Start date and time (UTC) \\
\hline
15149 & ACIS-I & 148.07 & Very Faint & 12 25 42.13 & +13 03 29.33 & 2013 Jul 24 04:22:25 \\
16967 & ACIS-I & 19.81 & Faint & 12 26 11.80 & +12 56 45.50 & 2016 May 2 06:37:07 \\
\hline
\end{tabular}
\end{table*}

We used a $148$-ks archival \textit{Chandra} Advanced CCD Imaging Spectrometer I (ACIS-I) observation from 2013 July, ObsID 15149, with PI Scott Randall. In 2016 May, prior to M86 tULX-1's discovery, we obtained a $20$-ks ACIS-I observation, ObsID 16967, aimed at the centre of M86. More details of these observations are listed in Table~\ref{tab:obs}.
We identified sources in these two observations with the CIAO tool \textsc{wavdetect} \citep{freeman2002}, with the \textsc{sigthresh} parameter set at $10^{-6}$, and the \textsc{scales} values a geometric progression with a constant factor of $\sqrt{2}$ between $2$ and $16$.

We extracted and modelled the \textit{Chandra} spectrum and (upper limit to the) flux of M86 tULX-1 using the CIAO software package.
We used a circular source region around M86 tULX-1 with radius of $4\farcs6$ and an annular background region with inner and outer radii of $4\farcs6$ and $23\farcs2$, respectively.
Neutral hydrogen in the interstellar medium absorbs part of the soft X-rays, which the spectral models take into account. The Galactic column density of neutral hydrogen atoms has an average value of $N_\mathrm{H} \approx 2.62 \times 10^{20}$ cm$^{-2}$ in the direction of M86, according to the HEASARC W3 $N_\mathrm{H}$ tool.
In order to avoid degeneracy between the spectral parameters, we froze the $N_\mathrm{H}$ parameter at this value in the spectral fits.

\section{Results}
\label{sect:ulx:results}

\subsection{Source location}

\begin{figure}
\includegraphics[width=\columnwidth]{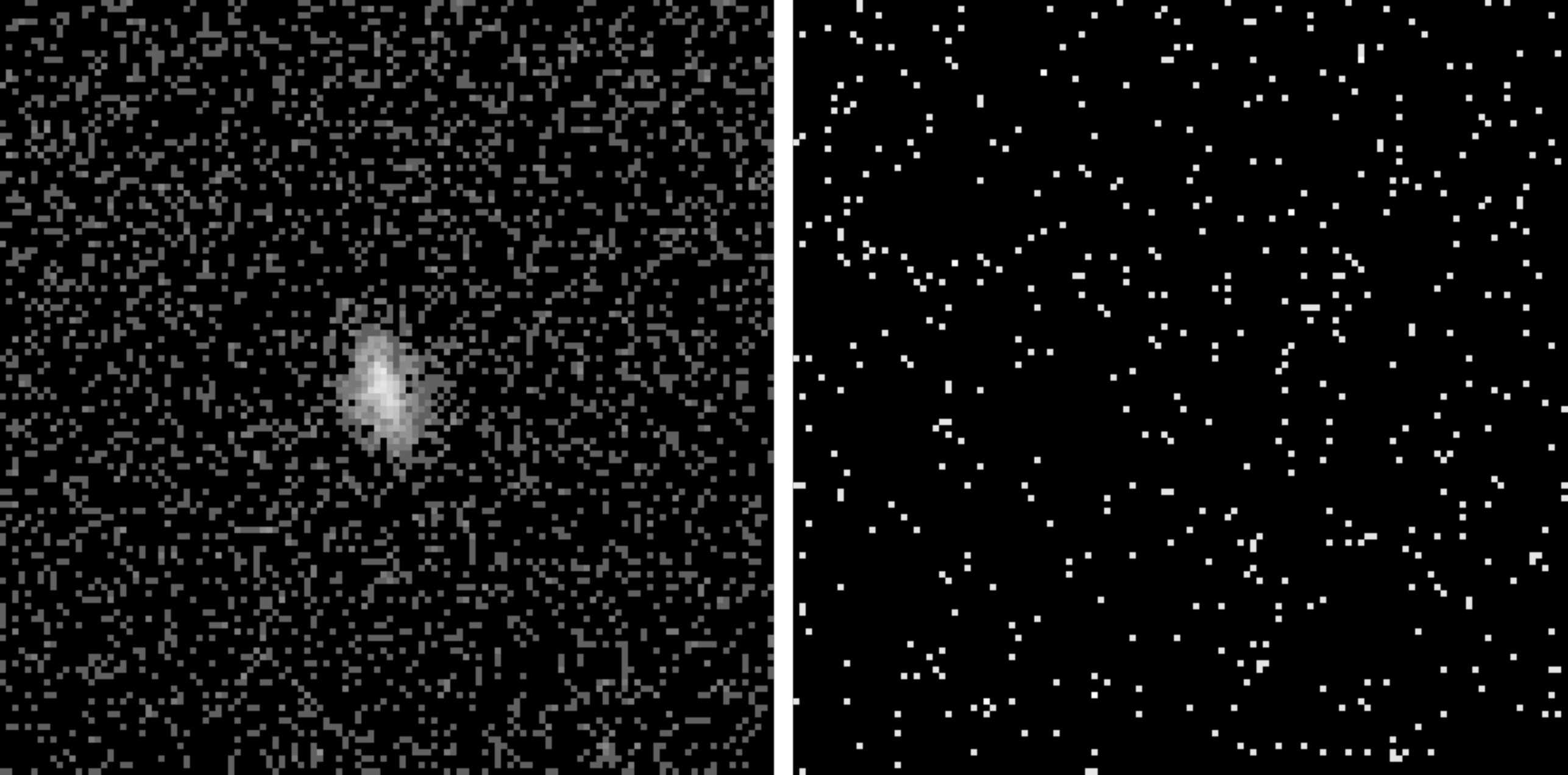}
\caption{\textit{Chandra} images of the $1 \times 1$ arcmin field centred around M86 tULX-1 on 2013 July 24--25 (left), and the same field on 2016 May 2 (right).}
\label{fig:ulx:skyimages}
\end{figure}

M86 is a member of the Virgo Cluster. The centre of the galaxy is located at RA 12$^{\mathrm{h}}$ 26$^{\mathrm{m}}$ 11\fs814, Dec. +12\degr 56\arcmin 45\farcs49 \citep{skrutskie2006}.
We adopted a distance to M86 of $16.83 \pm 0.54$ Mpc \citep{mei2007} to convert flux rates to luminosities.

We found M86 tULX-1 at RA 12$^{\mathrm{h}}$ 26$^{\mathrm{m}}$ 02\fs30, Dec. +12\degr 59\arcmin 51\farcs1 (J2000 coordinates) in ObsID 15149.
The projected distance of M86 tULX-1 to the centre of M86 is $3\farcm86$, which at the distance of the galaxy is equivalent to $19.3$ kpc. Since the effective radius of M86 is $3\farcm38 \pm 0\farcm24$ \citep{kormendy2009}, the galactocentric radius of M86 tULX-1 corresponds to $1.14$ effective radii.

The off-axis angle of M86 tULX-1 in ObsID 15149 is $6\farcm11$, and in ObsID 16967, in which this source is not detected, $3\farcm86$ (the same as the galactocentric distance, since the aimpoint of ObsID 16967 is at the centre of the galaxy).

The field around M86 tULX-1 in the two \textit{Chandra} observations is shown in Figure~\ref{fig:ulx:skyimages}.

\subsection{Light curve during bright epoch}

\begin{figure}
\includegraphics[width=\columnwidth]{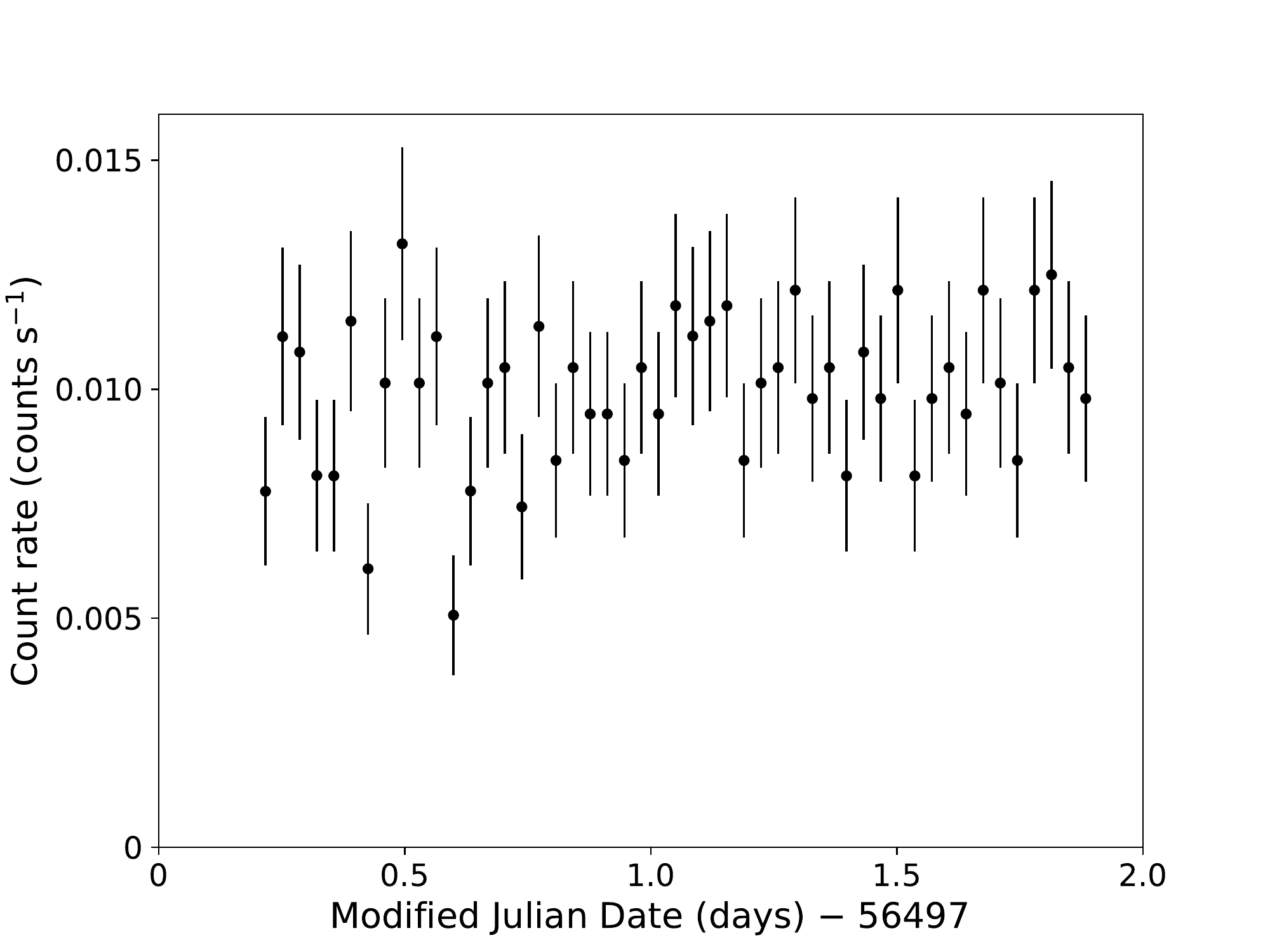}
\caption{Light curve of M86 tULX-1 during its bright epoch in 2013 July using 3000-second time bins, for the $0.5 - 8$ keV energy range. The average number of counts per bin is $29.5$.}
\label{fig:ulx:lightcurve}
\end{figure}

The light curve of M86 tULX-1 during the 41-hour exposure in 2013 July is shown in Figure~\ref{fig:ulx:lightcurve}. The normalized rms deviation is $0.18$ for $3000$-s bins for the $0.5 - 8$ keV energy range.
The normalized excess variance \citep{turner1999} of the light curve is $(-1 \pm 8) \times 10^{-3}$ for $3000$-s bins.
This insignificant excess suggests an absence of variability, which is confirmed by the Fourier power spectrum.

\subsection{X-ray spectrum and luminosity during bright epoch}

In the bright epoch, M86 tULX-1 had a background-subtracted count rate of $(8.64 \pm 0.25) \times 10^{-3}$ counts s$^{-1}$ ($1\sigma$ confidence interval) in ObsID 15149. The total X-ray photon count is $1.28 \times 10^{3}$.

We fitted the unbinned spectrum of M86 tULX-1 with an absorbed power-law model (\textsc{powlaw1d}), with an absorbed multicolour disc blackbody model (\textsc{diskbb}), and with a superposition of both models. The resulting \textsc{powlaw1d} power-law indices and \textsc{diskbb} inner disc temperatures are listed in Table~\ref{tab:spectrum}. A $\chi^{2}$ statistic with the Gehrels variance function has been used \citep{gehrels1986}, and the background has been subtracted.

\begin{table}
\caption{Spectral fits of M86 tULX-1 for the $0.5 - 8$ keV range. $T_\mathrm{in}$ is the inner disc temperature. The total flux and \textsc{diskbb} component flux are in units of $10^{-13}$ erg s$^{-1}$ cm$^{-2}$, which corresponds to $(3.4 \pm 0.2) \times 10^{39}$ erg s$^{-1}$ in M86 (the error is due to the uncertainty in the distance to M86). The \textsc{xspec} \citep{arnaud1996} goodness of fit of a model is the percentage of simulations based on that model with a test statistic less than that of the data. Confidence limits are $1\sigma$.}
\label{tab:spectrum}
\resizebox{\columnwidth}{!}{
\hspace*{-0.5cm}
\begin{tabular}{ccccccc}
\hline
Model & Power-law index & $T_\mathrm{in}$ (keV) & Flux & \textsc{diskbb} flux & Goodness  \\
\hline
\textsc{powlaw1d}        & $1.86 \pm 0.10$ &                 & $1.60 \pm 0.07$ & & $55\%$  \\
\textsc{diskbb}          &                 & $0.87 \pm 0.08$ & $0.98 \pm 0.04$ & & $90\%$ \\
\textsc{powlaw1d+diskbb} & $1.0^{+0.8}_{-2.6}$ & $0.66^{+0.17}_{-0.11}$ & $1.42^{+0.54}_{-0.18}$ & $0.6^{+0.9}_{-0.4}$ & $59\%$ \\
\hline
\end{tabular}
}
\end{table}

\begin{figure}
\includegraphics[width=\columnwidth]{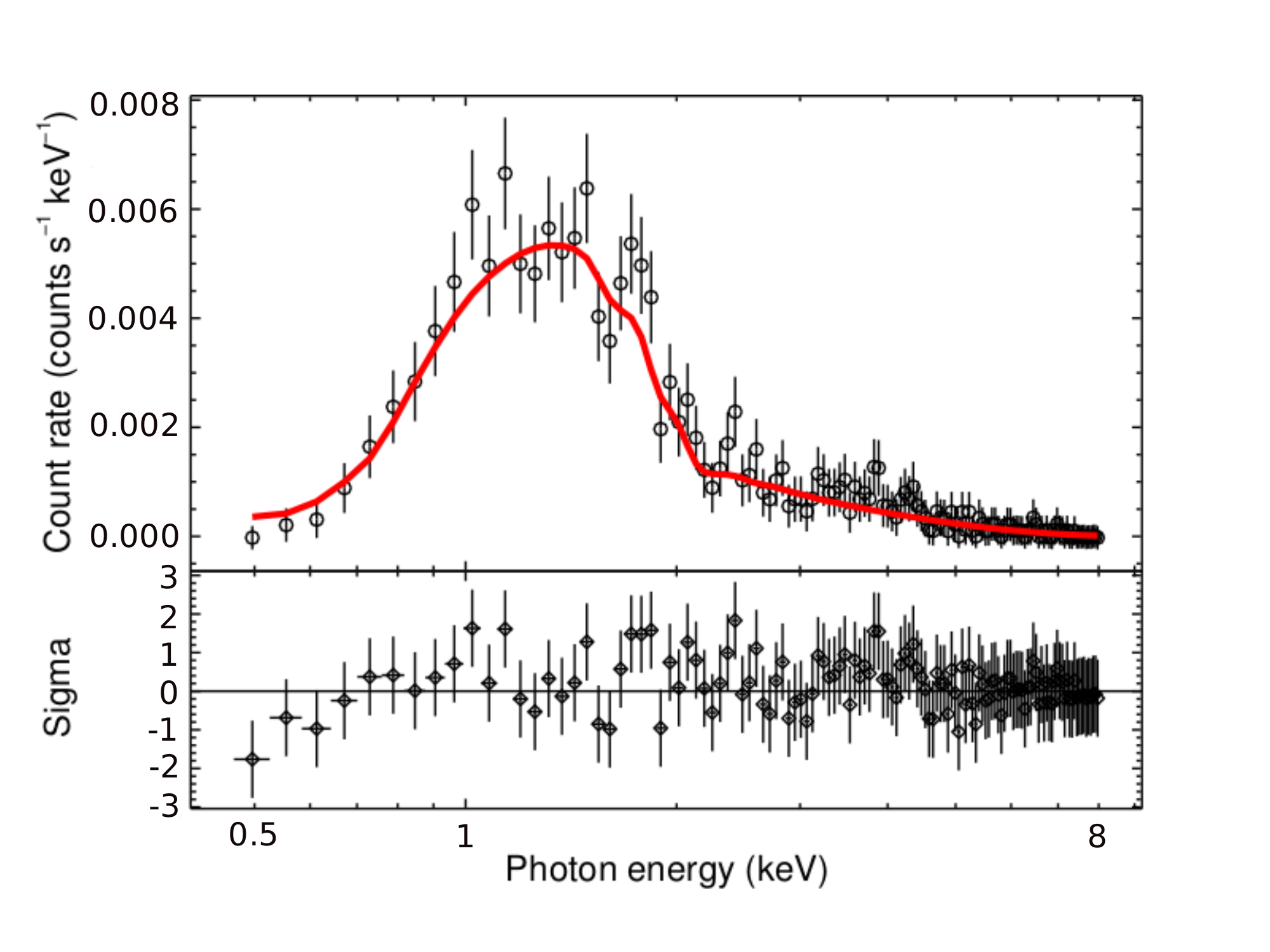}
\caption{Upper panel: X-ray spectrum of M86 tULX-1 for the $0.5 - 8$ keV range. The red curve shows the power-law plus disc blackbody fit with power-law index of $1.0$ and inner disc temperature of $0.66$ keV. Bottom panel: Residuals of the fit. The error bars are $1\sigma$.}
\label{fig:ulx:spectrum}
\end{figure}

The spectrum of M86 tULX-1 can be best fit (goodness of fit closest to $50\%$) by a power-law plus multicolour disc model with a $1.0^{+0.8}_{-2.6}$ index and a $0.66^{+0.17}_{-0.11}$ keV inner disc temperature, as shown in Figure~\ref{fig:ulx:spectrum}, or with an absorbed power-law model with index $1.86 \pm 0.10$. The large errors for the superposition model are the consequence of neither component model dominating -- a wide range of combinations of two components give a reasonably good fit. For an isotropically emitting source at the distance of M86, the luminosity based on this spectral model is ${(5 \pm 1) \times 10^{39}}$ erg s$^{-1}$, where the error is due to the error in flux rate, as well as due to the error in the distance to M86.

\subsection{Upper limit during faint epoch}

Using the CIAO tool \textsc{srcflux}, we determined the upper limit to the flux of an undetected source in ObsID 16967 using the point spread function area and location of the detection in ObsID 15149, when it was bright. The $3\sigma$ limit is $f_\mathrm{X} \approx 4.5 \times 10^{-15}$ erg s$^{-1}$ cm$^{-2}$ for the $0.5 - 8$ keV energy range, about $30$ times fainter than M86 tULX-1 during its bright epoch.

We checked archival \textit{XMM--Newton} data to see if M86 tULX-1 has been detected. Using the FLIX tool for \textit{XMM--Newton} upper limits by the XMM--Newton Survey Science Centre\footnote{https://www.ledas.ac.uk/flix/flix\_dr7.html} \citep{watson2001} at the location of M86 tULX-1, we found an upper limit of $f_\mathrm{X} = (1.3 \pm 0.3) \times 10^{-14}$ erg s$^{-1}$ cm$^{-2}$ in the $0.5 - 4.5$ keV band for a MOS-2 CCD observation (ID 0108260201) from 2002 July, and an upper limit of $f_\mathrm{X} = (3.7 \pm 4.6) \times 10^{-15}$ erg s$^{-1}$ cm$^{-2}$ for a pn CCD observation (ID 0673310101) from 2011 June, using a detection likelihood threshold of $10$, and a $5\arcsec$ radius. The source is not significantly detected in these observations, and we did not analyse them further.
The \textit{XMM--Newton} upper limit for the 2011 June observation is similar to the \textit{Chandra} value from 2016 May, and corresponds to $L_\mathrm{X} = (1.3 \pm 1.6) \times 10^{38}$ erg s$^{-1}$.

\subsection{Archival optical data}
\label{sect:optdata}

We used archival $B$-, $V$-, and $R$-band photometric data that were obtained with the Kitt Peak 4-m telescope and Mosaic camera for the purpose of studying the GC population of M86 \citep{rhode2004,young2016}.
The optical data at the location of M86 tULX-1 were taken on 1999 March 24, over 14 years earlier than the two X-ray observations we analysed.
The positions of the optical sources have typical $1\sigma$ errors of $0\farcs2$. The positional uncertainty of M86 tULX-1 is $0\farcs3$.
The nearest source to M86 tULX-1 is located at a distance of $13\farcs3$, so no optical counterpart associated with M86 tULX-1 is detected.
Several of the nearby optical sources have been identified based on their magnitudes and colours as GC candidates; the closest of these is located at $15\farcs0$ from M86 tULX-1.
The $50\%$ detection completeness limit for point sources is $m_V = 24.17$; GCs are usually brighter than this \citep{forbes1996}. Furthermore, because of the large distribution in GC masses, the percentage of GC stellar mass located in GCs fainter than this limit is significantly lower than the percentage of GCs fainter than this limit themselves. $1.9\%$ of GC mass is located in undetected GCs, based on the luminosity function for M86 in \citet{forbes1996}. Therefore, M86 tULX-1 is probably not located in a GC.

We also note that no sources within $10\arcsec$ in any wavelength band are listed in VizieR database of astronomical catalogues \citep*{vizier2000}.

\section{Discussion}
\label{sect:ulx:discussion}

\subsection{Comparison with other ULXs}

We highlight some of the unusual properties of M86 tULX-1 and its environment by comparing this source with other ULXs.

\subsubsection{Luminosity and environment}

The lack of ULXs in elliptical galaxies is consistent with a break near $L_\mathrm{X} = 5 \times 10^{38}$ erg s$^{-1}$ in the X-ray luminosity function of elliptical galaxies not affected by recent mergers \citep{sarazin2000,kim2010}.
ULXs with $L_\mathrm{X} \gtrsim 2 \times 10^{39}$ erg s$^{-1}$ are extremely rare in old populations.
The catalogue by \citet{swartz2011} lists two ULXs in the elliptical galaxy NGC 4150 with $L_\mathrm{X} \approx (5.6; 10.8) \times 10^{39}$ erg s$^{-1}$, but this galaxy has star-forming regions near these sources, especially near the brighter of the two.
Several bright ULXs have been found in GCs, e.g., in NGC 1399 at $L_\mathrm{X} = 5 \times 10^{39}$ erg s$^{-1}$ \citep{angelini2001,feng2006}, and a black-hole system in NGC 4472 \citep{maccarone2007}.
Overall, M86 tULX-1 may be the brightest ULX in an old field environment, unaffected by recent dynamical encounters.
M86 tULX-1 is the second ULX identified in M86, after CXO J122611.830+125647.80 at $L_\mathrm{X} = 2.2 \times 10^{39}$ erg s$^{-1}$ \citep{liu2011}.

\subsubsection{Spectral hardness}

\begin{figure}
\includegraphics[width=\columnwidth]{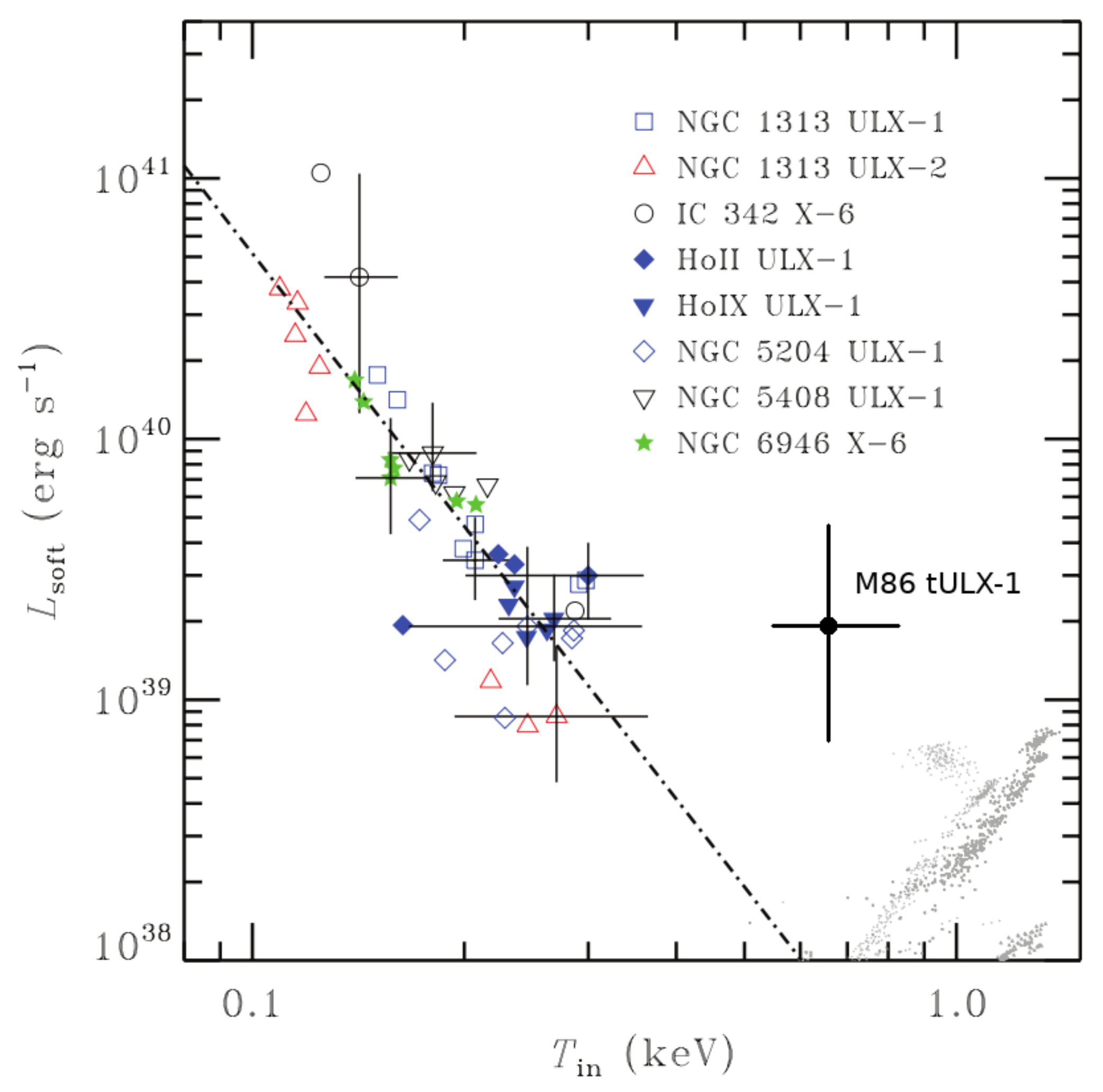}
\caption{Luminosity-temperature diagram of the soft excess of a sample of ULXs and Galactic X-ray binaries, reproduced with permission from \citet[][their figure~4]{kajava2009}. M86 tULX-1 has been added as a black dot.}
\label{fig:ulx:kajava}
\end{figure}

The inner disc temperature of $0.66$ keV is hotter than typical for ULXs, but also cooler than most Galactic black-hole X-ray binaries \citep{kajava2009}. Combined with its soft excess X-ray luminosity of $L_\mathrm{X} = (1.9 ^{+2.8}_{-1.3}) \times 10^{39}$ erg s$^{-1}$, M86 tULX-1 is located in an unpopulated region of the disc temperature-soft excess luminosity diagram \citep[][their figure~4]{kajava2009}, which we reproduced in our Figure~\ref{fig:ulx:kajava}.
The object thus is consistent with being a transitional object, between the bright ULXs and `normal' X-ray binaries.

\subsubsection{Variability}
\label{sect:varcompare}

ULXs in star-forming environments have been found to be more variable than those in old populations, which are mostly persistent \citep{feng2006,kaaret2017}.
For example, CXOU J132518.2$-$430304, a ULX in Centaurus A, became brighter by a factor of over $770$ in four years \citep{sivakoff2008}. CXOU J235808.7$-$323403, a ULX in NGC 7793, increased in luminosity from $L_\mathrm{X} < 5 \times 10^{37}$ erg s$^{-1}$ to $L_\mathrm{X} = 3.5 \times 10^{39}$ erg s$^{-1}$ (a factor of $>70$) in one year \citep{motch2014atel}, whereas NGC 5907 ULX-2 became brighter by a factor of over $35$ to reach $L_\mathrm{X} = 6.4 \times 10^{39}$ erg s$^{-1}$ \citep{pintore2018}.
All ULXs hosting NSs have been found to be variable by factors of ${\sim} 200$ \citep{furst2016}, and over $500$ \citep{israel2017sci, carpano2018}.

In old populations, several sources show outbursts on top of an already bright baseline. Short-duration flares have been observed up to ${\sim} 9 \times 10^{40}$ erg s$^{-1}$ in a GC source \citep{irwin2016}. XMMU 122939.7+075333, an $L_\mathrm{X} = 4 \times 10^{39}$ erg s$^{-1}$ black-hole X-ray binary in a GC, was found to fade to $L_\mathrm{X} = 1 \times 10^{38}$ erg s$^{-1}$, a factor of $40$ \citep{maccarone2010,joseph2015}.
CXO J122518.6+144545, either a hyperluminous X-ray source or a field ULX, varied by a factor $60$ \citep*{heida2015}.
Generally, ULXs in old (field) populations are rarely as variable as M86 tULX-1.

\subsection{Accretor type and mass}
\label{sect:accmass}

Based on its high luminosity of $L_\mathrm{X} \approx (3 - 6) \times 10^{39}$ erg s$^{-1}$ and its variability, M86 tULX-1 likely contains a BH. A $1.4\ \mathrm{M}_{\odot}$ NS accretor would imply an accretion rate of $34$ times the Eddington limit assuming hydrogen-rich matter, and $17$ times in case of helium or carbon/oxygen (assuming only electron scattering contributes to opacity).
Four ULXs hosting NSs with higher luminosities than M86 tULX-1 have been discovered \citep{bachetti2014,furst2016,israel2017mnras,israel2017sci,brightman2018} as well as one with similar luminosity \citep{carpano2018}. These may have very strong magnetic fields, allowing X-ray luminosities to be around $10^{40}$ erg s$^{-1}$ via accretion \citep{eksi2015,dallosso2015,mushtukov2015,tsygankov2016} or spin-down \citep{medvedev2013}, or alternatively, may have weaker magnetic fields of order $10^{12}$ G \citep{walton2018p13,walton2018ngc300}. High magnetic field NSs are short-lived \citep{thompson1995} and associated with supernova remnants and recent star formation \citep[e.g.,][]{kaspi2017}. In addition, the masses of companion stars to NS ULXs are observed or inferred to be over $5\ \mathrm{M}_{\odot}$ \citep[e.g.,][]{karino2018}. Located in an old environment, M86 tULX-1 is unlikely to be an accreting NS.

By simulating light curves that include a periodic signal, we ruled out at the $3\sigma$ level periodicities larger than $5.3$ s, for a pulsed fraction \citep[e.g.,][]{furst2016} of $50\%$, as well as periodicities larger than $11$ s for a pulsed fraction of $30\%$. The data quality and time resolution do not allow ruling out spin periods smaller than $2$ s, like those detected in three NS ULXs \citep{bachetti2014,furst2016,israel2017mnras,israel2017sci}, even for a pulsed fraction of $80\%$. A spin period of $31.6$ s, detected in NGC 300 ULX-1 \citep{carpano2018}, would only go undetected if the pulsed fraction is smaller than $24\%$, which is smaller than the values found by \citet{carpano2018}. In summary, neither an NS accretor with a ${\sim} 1$ s spin period, nor an NS with a long spin period but a very low pulsed fraction, can be ruled out for M86 tULX-1. However, an NS with the spin and pulse parameters of NGC 300 ULX-1 can be ruled out confidently.

M86 tULX-1 is among the most variable ULXs of similar or higher luminosity in which the presence of an NS has neither been established nor ruled out (see Section~\ref{sect:varcompare}). Its variability over a time-scale of at most three years could imply that it is accreting at a sub-Eddington rate \citep[based on the disc instability model, for a review see e.g.][]{lasota2001}, pointing to a massive BH if that is the case.
The relatively hard spectrum compared to other ULXs \citep[e.g.,][]{sutton2013,pintore2017,walton2018spectra} may be an indication of the low/hard state (but this is not necessarily the case), which corresponds to a luminosity of $\lesssim 0.03\ L_\mathrm{Edd}$ for Galactic BH low-mass X-ray binaries \citep{maccarone2003}.
If that fraction would also apply to M86 tULX-1, then its $L_\mathrm{Edd} \gtrsim 2 \times 10^{41}$ erg s$^{-1}$, i.e., the Eddington limit of a $\gtrsim 1500\ \mathrm{M}_{\odot}$ BH.
However, the inner disc temperature depends on the BH mass as $T_\mathrm{in} \propto M_\mathrm{BH}^{-1/4}$, implying a disc temperature of ${\sim} 0.1$ keV for accreting BHs that are ${\sim} 1500\ \mathrm{M}_{\odot}$ \citep*{winter2006}. This is much cooler than the $0.66$ keV we fitted for M86 tULX-1 in our combined power-law plus disc blackbody model, which better resembles a stellar-mass BH \citep{winter2006}.

For this reason, and because of the rarity of ULXs this luminous in elliptical galaxies, we consider a relative massive stellar-mass BH of ${\sim} 30 - 100\ \mathrm{M}_{\odot}$ to be a possible accretor in M86 tULX-1. Accretion at or below the Eddington limit would imply a $\gtrsim 48\ \mathrm{M}_{\odot}$ BH. A mass of several tens of solar masses is more massive than BHs in Galactic X-ray binaries whose masses have been measured dynamically, but similar to BH masses of around $30-60\ \mathrm{M}_{\odot}$ observed in the gravitational wave merger event GW150914 \citep{abbott2016gw150914}.
Alternatively, M86 ULX-1 could be in the transition region between the normal stellar-mass BHs and the UL state.

\subsection{Companion star}

The companion star is likely a giant, as donor stars with low average densities imply long binary orbital periods, and therefore high critical mass transfer rates \citep[e.g.,][]{zand2007,wu2010}, explaining the unstable accretion disc. Assuming a critical mass transfer rate of $\dot{M}_\mathrm{crit} \approx 5 \times 10^{-7}\ \mathrm{M}_{\odot}$yr$^{-1}$ (based on the peak X-ray luminosity), the orbital period is about $1 - 2$ weeks for the range of possibly realistic BH masses of $30 - 1500\ \mathrm{M}_{\odot}$ (see Section~\ref{sect:accmass}) \citep[][equation~1]{zand2007}. These orbital periods are near the high end of the range seen in Galactic low-mass X-ray binaries \citep*{liu2007}.
Any companion stellar type is too faint to be seen by ground-based observations.

\subsection{Ruling out alternative explanations}
\label{sect:ulx:alt}

Here we consider explanations for M86 tULX-1 other than a ULX in M86, both extragalactic and Galactic.

\subsubsection{Active galactic nucleus}

Active galactic nuclei (AGNs) with at least the X-ray flux of M86 tULX-1 occur at a rate of about 3 deg$^{-2}$ \citep{lamassa2016}, or ${\sim} 0.05$ in an area the size of M86. A chance coincidence therefore cannot be ruled out based on X-ray flux alone.

M86 tULX-1 faded to less than $3\%$ of its maximum observed X-ray luminosity within three years. AGNs are not this variable in X-rays over this timespan \citep[e.g.,][]{gibson2012,lanzuisi2014,yang2016}.

The optical data we used and the X-ray data during the bright epoch are not simultaneous, but because AGNs usually vary by less than one magnitude in the optical over several years \citep[e.g.,][]{giveon1999,helfand2001,kasliwal2015,zhang2017},\footnote{Except for BL Lac objects and related objects, but those are brighter than $m_{V} = 22$ \citep{kapanadze2013} and should have been detected in our deep optical images.} we assumed that M86 tULX-1 had $m_{V} \gtrsim 23$ during the X-ray bright phase.
M86 tULX-1 has a $0.5 - 2$ keV flux of ${\sim} (1.0 \pm 0.1) \times 10^{-13}$ erg s$^{-1}$ cm$^{-2}$ using an absorbed \textsc{powlaw1d+diskbb} model. Based on \citet[][their figure 11]{lamassa2016}, AGNs with fluxes of at least this value have $r$-band magnitudes brighter than $21$, while a few X-ray bright unidentified sources are as faint as $r = 22$. Two AGNs have a $0.5 -2$ keV flux of $(4 - 5) \times 10^{-14}$ erg s$^{-1}$ cm$^{-2}$ and $r \approx 23.5-25$, which is consistent with M86 tULX-1 within the error margins, but these two are clear outliers from the sample of ${\sim} 1700$ AGNs.

In summary, we conclude that M86 tULX-1 is too optically faint and too strongly variable in X-rays to be an AGN.

\subsubsection{Gravitational microlensing event}

An object in M86 could cause a microlensing event of an otherwise faint background AGN, leading to a bright X-ray transient.
For a range of possible lens masses of $10^{-3} - 1\ \mathrm{M}_{\odot}$, located in M86, and a source distance assumed to be much larger than the lens distance, we calculated how closely the lens and the source need to be aligned to produce a brightening of a given factor, using equations from \citet{paczynski1986}. The brightening factor must be at least $30$ for sources at the detection limit of our 2016 observation, or higher for fainter AGNs. We used the AGN luminosity function by \citet[][their figure 3]{lamassa2016} to calculate the number densities of AGNs available to produce such a brightening, for AGNs up to $3 \times 10^{-15}$ erg s$^{-1}$ cm$^{-2}$.

The expected number of AGN brightening events in the \textit{Chandra} field of view at any given time is less than $10^{-5}$, and unlikely to explain M86 tULX-1. The result is not sensitive to the lens mass or AGN lower limits.

\subsubsection{Tidal disruption event}

The disappearance of M86 tULX-1 in the 2016 \textit{Chandra} data is consistent with a tidal disruption event (TDE) of a star near a supermassive BH in a galaxy in the background of M86. For a typical X-ray luminosity of $L_\mathrm{X} \approx 10^{44}$ erg s$^{-1}$ \citep[e.g.,][]{vanvelzen2018}, the flare of a TDE with the flux of M86 tULX-1 would have a distance of a few Gpc, implying that the host galaxy would have an absolute magnitude fainter than $M_V \approx -18$. Such galaxies are bright enough to host supermassive BHs of order $10^{6}\ \mathrm{M}_{\odot}$, typical for TDEs \citep{auchettl2017,vanvelzen2018}.
However, the X-ray spectra of stars disrupted by BHs this massive are typically much softer than observed here \citep{auchettl2017}. For example, disc blackbody components with temperatures around $50$ eV have been measured in the cases of ASASSN-14li \citep{miller2015} and ASASSN-15oi \citep*{gezari2017}.

\subsubsection{Flare star in the Galactic halo}

The observed X-ray flux could potentially be caused by coronal X-ray emission from a young Galactic M- or K-dwarf in the foreground of M86, although a decay pattern would be expected over the $148$-ks light curve.
M86 tULX-1 is located at $15\degr$ from the Galactic North Pole, so this star would likely be located in the Galactic halo, rather than the thin or thick discs, unless it has a late M type.

The upper limit for X-ray luminosity $L_\mathrm{X}$ relative to bolometric luminosity $L_\mathrm{bol}$ for coronal flares from a wide range in late-type main-sequence stars is usually $L_\mathrm{X}/L_\mathrm{bol} \lesssim 10^{-2.5}$ \citep[][and references therein]{pizzolato2003,wright2011}.
For the X-ray flux of M86 tULX-1, it follows that the apparent magnitude $m_V$ is brighter than $18$, regardless of its distance.
Since the optical brightness of a main-sequence star is not expected to vary significantly over several decades, a foreground star close enough to Earth to produce the observed X-ray flux would easily have been detected by the optical survey.

\subsubsection{Very faint X-ray transient in the Galactic halo}
\label{sect:ulx:vfx}

At a distance of $3-30$ kpc in the Galactic halo, the flux of M86 tULX-1 corresponds to an X-ray luminosity of $L_\mathrm{X} \approx 10^{32-34}$ erg s$^{-1}$, which would make it a very faint X-ray binary \citep[e.g.,][]{degenaar2012fouryear}.
Given the Galactic latitude of ${\sim} 75\degr$, an X-ray binary in the Galactic thin disc would be located within a few $100$ pc and be brighter than the observed flux.

Based on the amount of stellar mass in the Galactic halo of ${\sim} 2 \times 10^{9}\ \mathrm{M}_{\odot}$ \citep[e.g.,][]{bullock2005}, and the theoretically modelled number of low-mass X-ray binaries of any luminosity per unit stellar mass of ${\sim} 10^{-7} \mathrm{M}_{\odot}^{-1}$ \citep[e.g.,][]{kiel2006,fragos2008,vanhaaften2015}, about $200$ hydrogen-rich X-ray binaries would be expected in the halo. A larger number of hydrogen-poor ultracompact X-ray binaries or their remnants could exist, but for old systems, low average mass transfer rates imply insufficient numbers of observable systems in outburst \citep{vanhaaften2013bulgeucxb}.
This corresponds to ${\lesssim} 10^{-3}$ in an area on the sky equal to the field of view of a \textit{Chandra} ACIS-I observation (if seen from the Galactic centre, less when seen from Earth), regardless of whether they are bright enough to be detected. However, M86 tULX-1 was not detected in 2016, and in general, many faint X-ray binaries could spend a large fraction of the time below $L_\mathrm{X} = 10^{32}$ erg s$^{-1}$ (if near $3$ kpc) and in particular below $L_\mathrm{X} = 10^{34}$ erg s$^{-1}$ (if near $30$ kpc), placing them below the detection limit of ObsID 15149. Therefore, the detection probability of a Galactic halo X-ray binary in a single \textit{Chandra} observation is a priori significantly lower than $10^{-3}$, but cannot be completely ruled out.

\subsubsection{Cataclysmic variable}

The X-ray flux of M86 tULX-1 is consistent with that of a faint cataclysmic variable (CV) in the Galactic thin disc at a distance of $100-300$ pc \citep{worrall1982}, or a brighter CV further out.

No source within $15\arcsec$ of M86 tULX-1 is detected by the GALEX \citep[Galaxy Evolution Explorer,][]{galex2005} All-Sky Imaging Survey in the NUV band ($1771-2831$~\AA), with a completeness limit of $22.5$ for the area around M86 tULX-1 \citep{bianchi2011properties}.
These data were taken several years before the M86 tULX-1 bright epoch, but the UV emission from a hot white dwarf is not expected to vary on such a short time-scale.

Non-magnetic CVs are rarely brighter than about $L_\mathrm{X} = 10^{32}$ erg s$^{-1}$ \citep{verbunt1997,mukai2017}, so non-magnetic CVs beyond a distance of $3$ kpc are a very unlikely explanation for M86 tULX-1.

The number of CVs in the halo with the observed flux of M86 tULX-1 per ACIS-I field of view can be estimated by integrating the space density of CVs multiplied by the stellar mass density along the line of sight towards M86 tULX-1. We used the space densities of non-magnetic and magnetic CVs by \citet{pretorius2012,pretorius2014} and the stellar mass density by \citet{bullock2005} to calculate a number less than $10^{-3}$ per ACIS-I field of view, using conservative assumptions.

\subsubsection{Supernova or gamma-ray burst afterglow}

A supernova in the elliptical galaxy M86 would be of Type Ia, and those are over $50$ times fainter in X-rays than M86 tULX-1 \citep{russell2012}. At the distance of M86, such an object should have been detected by optical transient searches.

A gamma-ray burst may have gone undetected, since the full sky is not continuously monitored. However, X-ray afterglows undergo clear decays on day time-scales, even \mbox{$1$ Ms} after the burst \citep{nousek2006,liang2007}, and this is not seen in the light curve in Figure~\ref{fig:ulx:lightcurve}.

\section{Conclusion}

The properties of M86 tULX-1 -- namely, its high luminosity, location in the field of an elliptical galaxy, relatively hard spectrum, and transient behaviour -- make this object an unusual ULX. Based on our analysis of both \textit{Chandra} data and ground-based optical imaging, we conclude that it may be a relatively massive stellar-mass black hole accreting from a low-mass giant companion, or a stellar-mass black hole in a transitional state between the normal stellar-mass black holes and those in the ultraluminous state.

Follow-up observations in various bands will potentially be valuable. Of particular use would be an \textit{XMM--Newton} observation during a bright epoch; this would provide a higher-quality spectrum to more accurately determine the object's spectral state (especially if combined with a simultaneous radio observation). Alternatively, obtaining \textit{XMM--Newton} data during a faint epoch would provide more information about the low-hard state. The latter spectrum could be compared to spectra of Galactic X-ray binaries with dynamically measured stellar-mass BH accretors to estimate the accretor mass in M86 tULX-1. A simultaneous observation by the planned \textit{James Webb Space Telescope} would be able to better constrain the nature of the system, aided by the system's location in a low surface brightness part of M86.

\section*{Acknowledgements}

LMvH thanks Paul~H. Sell for discussion of \textit{Chandra} data analysis, Lee Burnside for computer help, and the anonymous referee for valuable suggestions.
Support for this work was provided by National Aeronautics and Space Administration (NASA) through Chandra Award Number GO4-15034 issued by the Chandra X-ray Center (CXC), which is operated by the Smithsonian Astrophysical Observatory for and on behalf of NASA under contract NAS8-03060, and by NASA Astrophysics Data Analysis grant NNX15AI71G.
The scientific results reported in this article are based on observations made by the \textit{Chandra X-ray Observatory}, and on data obtained from the Chandra Data Archive.
This research has made use of software provided by the CXC in the application package CIAO, as well as NASA's Astrophysics Data System Bibliographic Services.

\bibliographystyle{mnras}
\bibliography{lennart_refs}

\bsp
\label{lastpage}
\end{document}